\documentclass{article}

\usepackage{amsmath}
\usepackage{PRIMEarxiv}
\usepackage{float}
\usepackage[utf8]{inputenc} % allow utf-8 input
\usepackage[T1]{fontenc}    % use 8-bit T1 fonts
\usepackage{hyperref}       % hyperlinks
\usepackage{url}            % simple URL typesetting
\usepackage{booktabs}       % professional-quality tables
\usepackage{amsfonts}       % blackboard math symbols
\usepackage{nicefrac}       % compact symbols for 1/2, etc.
\usepackage{microtype}      % microtypography
\usepackage{lipsum}
\usepackage{fancyhdr}       % header
\usepackage{graphicx}       % graphics
\graphicspath{{media/}}     % organize your images and other figures under media/ folder
\usepackage{hyperref}
\title{AlphaZip : Neural Network-Enhanced \\
Lossless Text Compression\\
\large A word to the wise is enough
}
\author{
  Swathi Shree Narashiman \\
  Undergraduate, Department of Electrical Engineering \\
  Indian Institute of Technology, Madras\\
  \texttt{rudhranarashiman.225@gmail} \\
   \And
  Nitin Chandrachoodan \\
  Professor, Department of Electrical Engineering \\
  Indian Institute of Technology, Madras \\
  \texttt{nitin@ee.iitm.ac.in} \\
}

\begin{document}
\maketitle
\begin{abstract}

Data compression continues to evolve, with traditional information theory methods being widely used for compressing text, images, and videos. Recently, there has been growing interest in leveraging Generative AI for predictive compression techniques. This paper  \footnote{GitHub Repository: \href{https://github.com/Swathi-Shree-Narashiman/AlphaZip}{AlphaZip}} introduces a lossless text compression approach using a Large Language Model (LLM). The method involves two key steps: first, prediction using a dense neural network architecture, such as a transformer block; second, compressing the predicted ranks \cite{stanford-report-2024} with standard compression algorithms like Adaptive Huffman, LZ77, or Gzip. Extensive analysis and benchmarking against conventional information-theoretic baselines demonstrate that neural compression offers improved performance.

\end{abstract}

% keywords can be removed

\section{Introduction}

Data compression, especially in the realm of text, is the art of reducing the size of information without sacrificing its integrity. In an era where vast amounts of data are constantly exchanged, efficient text compression has become more critical than ever, enabling faster communication, reduced storage costs, and enhanced performance in bandwidth-limited environments. Text in digital form is stored using different character sets (ASCII, Unicode, etc.) that are encoded into binary using encoding schemes like ASCII encoding, UTF-8, UTF-16 etc. Text files can be stored in plain text format devoid of any formatting metadata or could be stored in Rich Text, HTML or XML formats where structuring and formatting add an overhead to the storage space. Compressing text files involves identifying the redundancies in these representations and encoding them to capture recurring pattern information.  This is usually achieved through lossless compression algorithms. Though there have been many purely information theoretic frameworks for compressing text in a lossless manner, Neural Network based architectures outperform most of these techniques, as they can encode additional information about the relationships between different components of text\cite{bellard2021nncp}. 

Generative pre-trained transformers (GPTs) are the state-of-the-art in text generation and prediction~\cite{10500411}.  In this study, we leverage the power of Large Language models in our generative compression mechanism.  Our model applies a compression algorithm over the outputs from the transformer block and compresses the text to a binary bit stream that can be transmitted \cite{platos2008word}. Although the use of GPTs is computationally heavy, we show that by utilising accelerated linear algebra (XLA) compilation combined with an optimal neural model size the process can be performed in reasonable time for realistic data. 

We also present a behavioral analysis of various models on compression performance and  process latency. We show that by constraining the model using fine-tuning techniques, we can achieve domain based compression that performs better than agnostic techniques that do not use this information.

Information theoretic compression is well studied, and we do not expect any significant changes in the compression ratios achieved through standard compression algorithms.  Instead, the main focus of this paper is to see if the prediction capabilities of neural networks introduce additional redundancy that can then be exploited by standard algorithms.  With this in mind, we formulate generative compression as a two step process :
\begin{enumerate}
    \item Prediction of ranks using a Neural Network
    \item Compression of the predicted rank sequence using standard compression algorithms.
\end{enumerate}

Experimental analysis demonstrates that neural network-based predictive compression achieves an approximate 57\% improvement in compression ratio compared to a baseline based on the GZIP compression algorithm.  We suggest that this enhancement can primarily be attributed to the predictive capabilities of the large language model, indicating that the performance of the compression is closely linked to the accuracy of the predictor. Improved predictor performance consistently results in better compression ratios.

\section{Background}
Prior art in the area of text compression primarily points to two major text compression methods, namely entropy encoding and dictionary encoding. A purely probabilistic model based on greedy prediction, followed by compression analysis is proposed in Ref.~\cite{valmeekam2023llmzip}. Their work gives an understanding of how rank prediction can be exploited to capture the redundancies and compress text effectively. Another study\cite{cleary1984data}, tries to avoid the latency in two passes by proposing a one-pass adaptive method . This method models the chances of a character occurring in the same pre-context and accommodates the probability as the input is parsed in. DeepZip~\cite{goyal2018deepzip} utilises the power of Recurrent Neural Networks for compression and seems to outperform traditional compressors like GZIP \cite{gzip} and BSC \cite{binary_symmetric_channel}. Deepzip combines RNNs with arithmetic coder for compression making it effective compared to other finite context models. 

The Transformer architecture \cite{vaswani2017attention} has proposed the concept of \emph{attention} mechanisms as a useful way of predicting text.  The query, key, value based prediction in transformers surpassed all the state-of-the-art sequence to sequence models.  This was followed by many attempts including the usage of LSTMs \cite{bellard2021nncp} and transformers, that have shown interesting results. A recent work \cite{delétang2024languagemodelingcompression} emphasises the potential of Large Language Models (LLMs) in prediction, and advocates viewing the prediction problem through the lens of compression. It demonstrates how a compressor can be transformed into a predictor by developing probability distributions using Shannon's entropy principle. 

Tokenization can be viewed as a pre-compression step allowing models to increase the information content in their context. This drives our motivation to analyse the power of LLMs in lossless text compression in a rank based compression mechanism.

Fine-tuning a pre-trained model can improve the performance in task specific applications. To reduce the compute intensity PEFT\footnote{Parameter Efficient Fine-tuning Techniques} \cite{sabry2023peft} techniques like LoRA\footnote{Low Rank Adapatation} \cite{hu2021lora} are found to be helpful. Fine-tuned models are domain specific, reducing the uncertainty in prediction,  hence making the model a better compressor (although they cannot be generalised). This intuition has been studied in our experiments to compare the performance of domain specific LLMs in text compression. 

In contrast to standard information-theoretic compressors (which model the frequency of symbols for encoding) neural network-based predictors also capture the semantic similarity within a continuous stream of tokens. This approach effectively reduces the amount of information to be transmitted, thereby enhancing compression efficiency. Unlike many previous approaches that use probabilistic models to process the input text before compression, our experiments are conducted on unseen text (not present in the training data for fine tuning). As a result, the effectiveness of compression depends heavily on the accuracy of the neural network predictor.

\subsection*{Contributions} 
In this study, we conducted an experimental analysis on the compressive capabilities of large language models.  It is important to note that we have taken the approach of using the LLM for predicting tokens, and leaving the actual compression to a known algorithm -- this allows us to focus on the effectiveness of the LLM in predicting text.
\begin{itemize}
\item We examine how the predictive performance of the neural network can improve compression ratio.
\item We achieved a compression ratio of up to 57\% through the proposed two step predictive compression mechanism. Constraining the domain knowledge of the LLM proves to enhance the compression performance.
\item We show that smaller models like GPT2-small (124M parameters)~\cite{openai_gpt2} can compress text effectively through knowledge distillation from a suitable teacher model, thus eliminating the need for compute intensive models to solve the compression problem.
\item We analyse the compressive performance and process latency by varying the model size and parameter count. We also compare the performance of GZIP and Brotli \cite{wikipedia_brotli} in predictive compression.
\end{itemize}

\subsection{Information Theoretic Baselines}

Lossless compression algorithms reduce the redundancies in the data to ensure that there is no data loss when de-compressing data back to the original form\cite{book}.  Certain types of data such as images can survive a certain amount of loss in accuracy, while text compression comes under the lossless type\cite{article}. To quantify the performance of neural compression, we compare the results with few of the baselines discussed below.

\subsubsection{Arithmetic Encoding }

Arithmetic encoding is an entropy encoding that requires two inputs: the next symbol and its frequency. The frequencies are utilised to calculate the probabilities of each symbol. The result of arithmetic encoding is a single number ranging between 0 to 1, a variable length representation. This is a compact representation and requires further processing to convert it into a binary representation.

\subsubsection{Huffman and Adaptive Huffman Encoding }

Huffman encoding assigns variable length codes to symbols based on their frequency of occurrence. The probability distribution is used to construct a Huffman Tree using a bottom-to-top approach. The tree is traversed from the root to the leaf node to encode a given symbol.
\footnote{In our experiments, we use the standard Python module, dahuffman 0.4.1\cite{dahuffman_pypi} provided by PyPi.org.}
%%% FIXME: I don't see dahuffman as a standard module.  What is this? - Nitin: It is developed by PyPi which is a trusted organization sir, will change the footnotes. 
\footnote{
Our implementation of the Adaptive Huffman encoding adapted an existing implementation in GitHub. \hyperlink{https://github.com/A-s-h-w-i-n/Adaptive_Huffman/blob/master/adaptive_huffman.py}{link}}.

\subsubsection{Lempel Ziv 77 (LZ77) }

To capture the higher order relationships in words and phrases, Jacob Ziv and Abraham Lempel developed a sliding window concept. The standard implementation of the LZW (Lempel Ziv Welch) algorithm utilises a table (dictionary) of 4096 entries of symbols and their binary representation. The first 255 entries correspond to ASCII. As the input is parsed, the longest match is obtained from the dictionary (if found) else new entries are filled in the dictionary.

In contrast, the LZ77 algorithm does not store a dictionary: instead it uses a sliding window buffer to identify and encode repeated sub-strings by specifying their distance and length from the current position in the buffer. 

\subsubsection{Gzip compression algorithm }

Gzip \footnote{Utilised gzip compression module provided by the python.org. at compresslevel 9 (best compression standard).}
uses the DEFLATE compression algorithm\cite{gzip_wikipedia},
%%% FIXME: Cite details like this - Fixed
which combines LZ77 (Lempel-Ziv 1977) and Huffman coding techniques. LZ77 is used to find repeated sequences in the data, while Huffman coding is employed to assign variable-length codes to these sequences based on their frequencies.

\begin{enumerate}
    \item  Repeated sequences are identified and given shorter codes (LZ77).
    \item The codes with high frequencies are assigned fewer bits whereas the ones appearing rarely are assigned more number of bits (Huffman).
\end{enumerate}

\subsubsection{Brotli compression algorithm }

Brotli \footnote{ Brotli open source compression module provided by the python.org. at level 11 (best compression standard)} , a dictionary based compression algorithm  developed by Google \cite{47824} has been found to surpass most of the state-of-the-art compression techniques including gzip. This is attributed to  Brotli's context modeling method to adapt to the input data pattern.   

Brotli compression occurs in two phases: in the first phase, parsing based on the previous occurrences is used to create the static LZ77 dictionary, while in the second phase a Huffman encoding of the parsed data is performed by choosing the optimal encoding from the canonical Huffman forms. Thus by using a combination of static dictionary encoding and building dynamic Huffman trees, Brotli achieves better and faster compression than most of the LZ family counterparts.

\section{Neural Compression Model Architecture}
As discussed earlier, compression involves identifying redundancy in the input data and then efficiently encoding that redundancy.  In this work, we focus on mechanisms that can enhance the former part: namely using the predictive powers of LLMs to identify possible redundancy in the textual data, thereby providing added redundancy in the input that is provided to the standard compression algorithms.  In particular, we focus on the use of \textbf{transformer} based architectures to predict text in such a way as to enhance the capability of compression algorithms.

\vspace{0.2cm}

\begin{figure}[htbp]
     \begin{center}
     \includegraphics[width=0.5\textwidth]{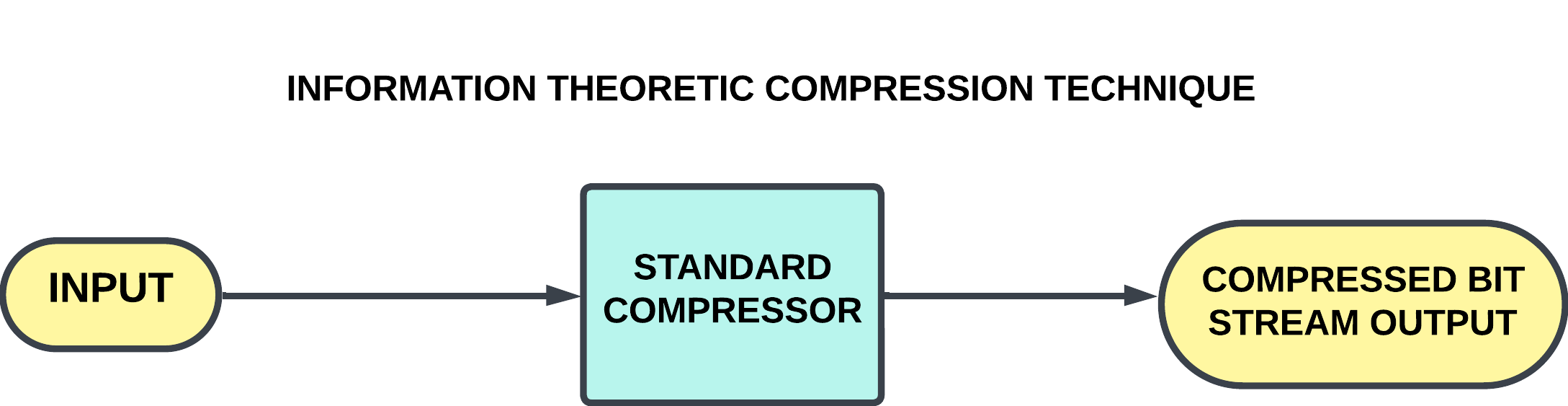}
     \end{center}
     \caption{Block diagram representing standard compression pipeline }
 \end{figure}
\vspace{0.6cm}

\begin{center}
 \includegraphics[width=0.65\textwidth]{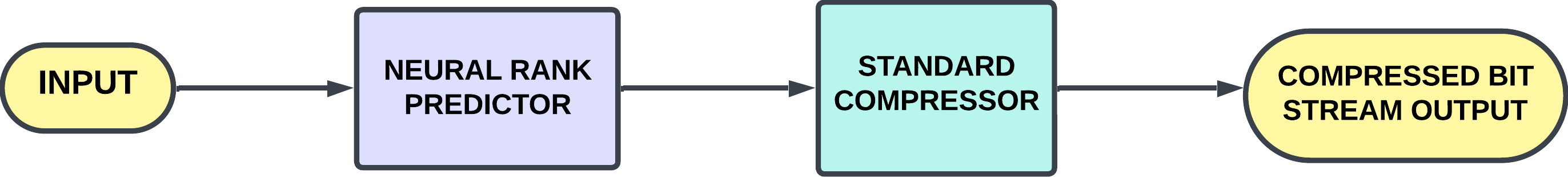}
\end{center}

\begin{figure}[htbp]
     \begin{center}
     \includegraphics[width=\textwidth]{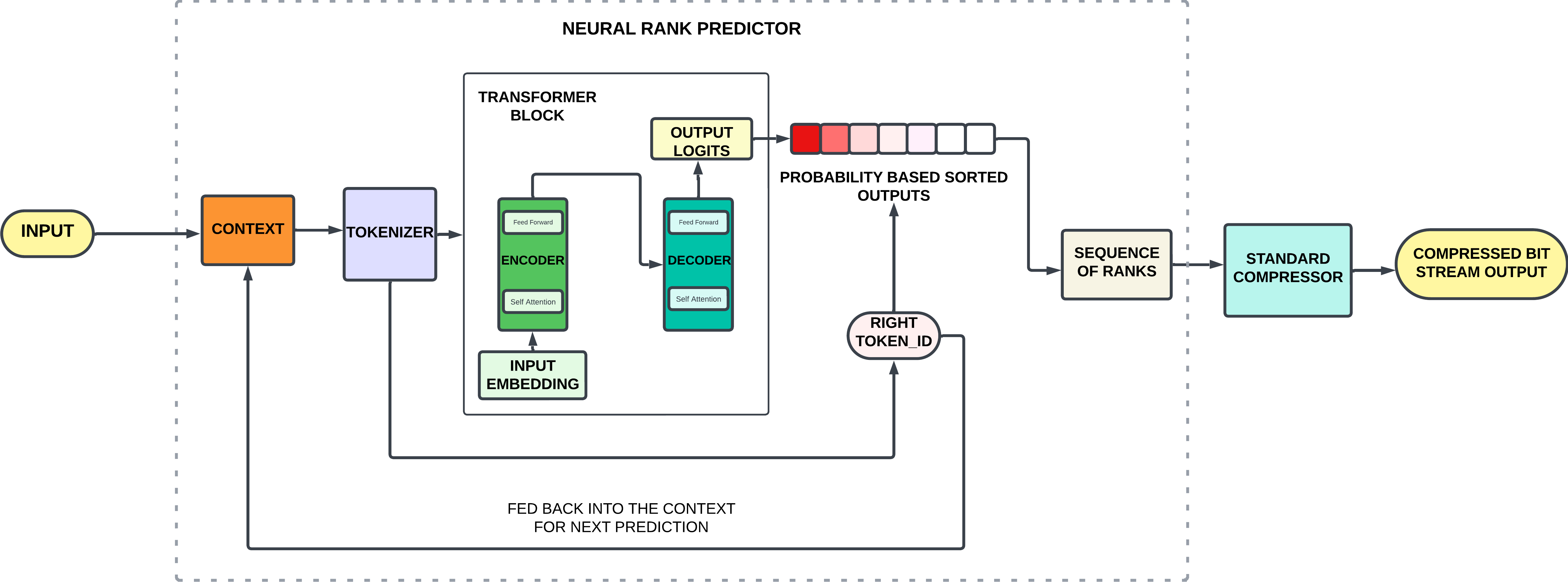} 
     \end{center}
     \caption{Block diagram representing the high level architecture of AlphaZip}
 \end{figure}

\subsection{Tokenization}
Tokenization is the process of breaking down an input text into smaller chunks that can be processed by Large Language models. GPT2\cite{openai_gpt2} uses Byte pair encoding\cite{huggingface_nlp_chapter6_5}. In this method, initially, each character is treated as a symbol. The algorithm then iteratively merges frequently occurring pairs of symbols into new symbols and adds them to the vocabulary. The final vocabulary converges to a point where the indices of each symbol represent its token ID. GPT2 also performs word-level encoding and text pre-processing to improve the model performance.

The input text is parsed through the tokenizer to obtain a tokenized vector. This vector is utilised to find the right token ID following a given context, and to predict the rank of the right token.

\subsection{Rank Prediction}

The rank of the right Token ID following a context is the heart of AlphaZip's architecture.  When a given context is fed into the Pre-Trained Transformer, the output logits~\footnote{Log-odds function - commonly used as a measure of predictive accuracy} are used to calculate the probability distribution across the entire vocabulary of tokens in the tokenizer. By comparing it with the tokenized input, we calculate the rank of the right token ID based on the sorted probability distribution. This means that the most probable token being the right token corresponds to rank 0, the second most probable token being the right token corresponds to rank 1 and so on. 

It is observed that the distribution of these ranks decreases with the order of the rank with 0 being the most recurring rank. The variation of the distribution of ranks \textit{vs} the rank plotted below clearly shows a curve that has an \emph{exponential decay} behavior. The better the predictor, ranks close to 0 occur more frequently, thus improving the compression ratio. The sequence of ranks encode the redundancy in the input string while the actual representation of these ranks has little effect on the compression. Hence, we feed the ranks separated by '.'s as ASCII into the standard compressor.  It is quite possible that better compression algorithms could be identified to further compress this data -- however, our goal here is to identify redundancy, and hence we fall back to traditional lossless compression algorithms to remove the redundancy in the data.

\begin{figure}[htbp]
     \begin{center}
     \includegraphics[width= 0.7\textwidth]{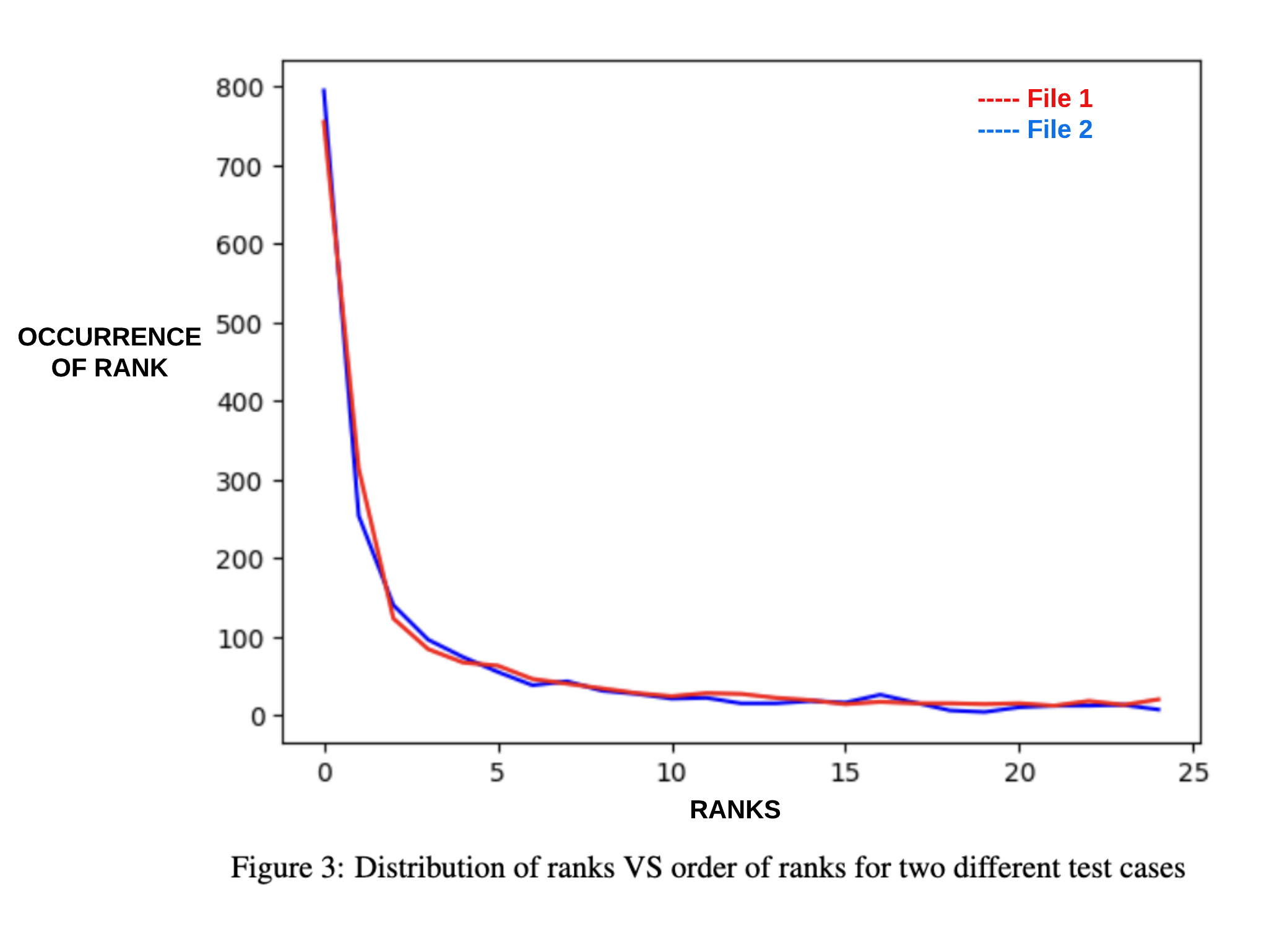} 
     \end{center}
     \caption{Distribution of ranks \textit{vs} order of ranks for two different test cases.}
 \end{figure}

\subsection{Individual Inference \textit{vs} Batch Inference}

Inferencing is the process of applying a given model at the individual token level to infer the predicted values corresponding to a given input.  This  involves computing probabilities for every token in the tokenized input given the right previous context. This increases the latency and time of computing. Batch inferencing on the other hand computes the output probabilities for the next \textbf{k} tokens given a context size.

While individual inferencing follows a greedy prediction algorithm, batch inferencing introduces some stochasticity in the model prediction and hence achieving ranks closer to 0 becomes less likely. Optimising the value of \textbf{k} to get better results can help set the trade off between latency and compression performance. In all the experiments that follow, we have opted to utilise Individual inferencing.

\subsection{Accelerating inferencing using TensorFlow XLA}

TensorFlow's XLA (Accelerated Linear Algebra) \cite{openxla_xla} is a domain specific compiler that optimizes the execution of computation graphs especially for matrix multiplications and vector operations and hence increases performance. Studies show that XLA can show a 50\% speed up in computation in GPUs over TensorFlow models without XLA\cite{openxla_xla}.

During XLA compilation, the initial computation graph is optimized through several passes such as Dead Code Elimination and Common Sub-expression Elimination~\cite{snider2023operator}. The fusion techniques applied include horizontal fusion, instruction fusion, multi-output fusion etc. In our implementation procedures, we use Just In Time (JIT) Compilation, during which frequently executed code paths are identified and compiled to machine code for better performance. JIT compilation promotes adaptability during compilation and eliminates startup time required in Ahead of Time (AOT) compilation. 
% At best there could be some minimal JIT passes - not continuous improvements - Agreeing on this sir, I am removing this claim
However, XLA shows poor performance when the input tensor has varying shape, compelling us to maintain a fixed context window size.

\begin{table}[H]
  \centering
  \begin{tabular}{llll}
    \toprule                  
    Model    & Size &  With XLA (s)     & Without XLA  \\
    \midrule
    gpt2 & 124 M &70.94 &    $\approx$ 15 minutes  \\
    gpt2-medium  & 355 M  & 147.44 & $\approx$ 30 minutes     \\
    gpt2-large  &  774 M  &  229.79   &  $\approx$ 1 hour  \\
    \bottomrule
  \end{tabular}
  \vspace{0.1cm}
   \caption{Time taken for compression (averaged) for different test cases of the same size}
\end{table}

\section{Results and Analysis}

We conduct a set of experiments to compress works of various authors.  We quantify compression using bpc and compression ratio metrics. To maintain uniformity and ease out comparative analysis we compress only the first 10000 characters in every book. We use TensorFlow XLA during compression  throughout the experiments. The obtained ranks from the transformer architecture are compressed using the gzip algorithm to get the output bit stream. We analyse the style transfer capability of transformer based architectures in domain based compression. Through results, we state that domain specific transformers perform better compression than vanilla transformer models.

\subsection*{Datasets and Training Arguments}

We  utilise the open source Gutenberg corpus \cite{lahiri:2014:SRW} for all our tests and analysis. The corpus has books stored in .txt format and does not require any pre-processing.  To compress multi-lingual text we utilised resources from the OPUS dataset\cite{tiedemann2012parallel}. For a few experiments, we would also be using the Tiny Shakespeare dataset \cite{karpathy2015char_rnn}. 

 \subsection*{Hyperparameters}
To show our compression results and performance, we try to compress nearly the first 10KB (10000 characters) of data from each book in the corpus with the context window being of size 100. Through our experiments we observed that, bigger the context window size is longer the process latency is, thus choosing an optimal window size can set the trade off between accuracy of prediction and latency. For better convergence, we fine-tuned or knowledge distilled vanilla GPT2 models with a learning rate of 5e-7. 

\subsection{Compression Performance Quantification}

\textbf{Compression Ratio} is defined as the ratio of the uncompressed file size to the compressed file size.  Generally, a larger value of compression ratio is desirable.

\begin{equation}
Compression\ ratio = \frac{Uncompressed\ file\ size}{Compressed\ file\ size}
\end{equation}

\textbf{Bits Per Character (bpc)} is the ratio of the size of the compressed file in bits to the number of ASCII characters in the content of the original file. This is a popular metric for text, for other types of data bits per byte (bpb) is used. 

\begin{equation}
bpc = \frac{Compressed\ file\ size\ in\ bits}{Number\ of\ characters\ in\ original\ file}
\end{equation}

\textbf{Entropy} (Shannon Entropy) of a file is a measure of the average information content \cite{4068914} in the data source. Lower entropy indicates more predictable data, which can be compressed more effectively.

\begin{equation}
H = -\sum_{i=1}^{n} p_i \log_2(p_i)
\end{equation}
Where:
\\n is the number of unique symbols (bytes, characters, etc.) in the file.
\\ $p_i$ \text{  is the probability of occurrence of the } $i^{th}$ \text{ symbol in the file.} 

\textbf{Compression Time} is the time taken to compress the file and is an inherent measure of the compression speed. \footnote{In our experiments, we use the Python \texttt{timeit} library from the Python standard library.}

\subsection{Results on Vanilla GPT2}

Results of compression of the first 10000 characters of 8 works per author for different authors  are presented below and compared against the GZIP baseline. The context length is 100 characters. The compression ratio for the Gzip baseline is approximately 2, whereas that of the neural predictive compression model based on gpt2 (124 M) is close to 3.5 which is roughly 1.57 times that of GZIP. 

\begin{table}[H]
  \centering
  \begin{tabular}{llll}
    \toprule
    Dataset     & Gzip ratio & GPT2+GZIP ratio & average bpc \\
    \midrule
    Lord Byron & 2.16 & 3.01 & 2.76 \\
    W.B. Yeats & 2.28 & 3.33 & 2.41 \\
    Abraham Lincoln & 2.26 & 3.98 & 2.02 \\
    Winston Churchill & 2.09 & 3.62 & 2.22 \\
    average & 2.20 & 3.47 & 2.35 \\
    \bottomrule
  \end{tabular}
  \vspace{0.1cm}
  \caption{Results for compression on raw gpt2}
  \label{tab:table}
\end{table}

\subsection{Traditional fine-tuned GPT2 on a domain corpora}

Fine-tuning the model entails adjusting its weights and biases to optimize performance for a specific task using a targeted dataset. In this study, we fine-tune a raw GPT-2 model on a corpus consisting of 15 books by a single author, using a learning rate of 5e-7. We then evaluate the model's compression performance on texts by both the same author and different authors to assess its generalization capability.

Fine-tuned GPT2 performs better than raw GPT2 and improves compression by nearly 8\%. While compression also improves by at least 3\% for works of other authors. This suggests that some degree of style adaptation may occur during fine-tuning. Given that GPT-2 was originally trained on a diverse dataset of 8 million web pages spanning various genres, fine-tuning it on storybooks appears to have enhanced its ability to capture patterns specific to narrative writing. As a result, this may have contributed to improved compression performance in out-of-distribution (OOD) inference, though further analysis is needed to fully understand the impact.

\begin{table}[H]
  \centering
  \begin{tabular}{llll}
    \toprule
     Author &   W.B.Yeats(original) & Abraham Lincoln & Winston Churchill\\
    \midrule
     GZIP  ratio  &       2.28        &  2.26             &      2.09           \\
     GPT2+GZIP ratio   &       3.33      &    3.98         &    3.62           \\
     Fine-tuned GPT2 +GZIP &    3.61       &  4.17            &        3.81        \\
     \% improvement &     8.39       &  4.80       &       5.35         \\
    \bottomrule
  \end{tabular}
  \vspace{0.1cm}
  \caption{Results of compression on fine-tuned GPT2 on works of W.B.Yeats (checkpoint-8000)}
  \label{tab:table}
\end{table}

\begin{table}[H]
  \centering
  \begin{tabular}{llll}
    \toprule
     Author &   A.C. Doyle(original) & Zane Gray & Rudyard Kipling\\
    \midrule
     GZIP  ratio  &       2.10        &  2.17                &     2.28          \\
     GPT2+GZIP  ratio   &       3.68          &     3.47             &   3.15            \\
     Fine-tuned GPT2 +GZIP &       4.01    &   3.69              &        3.32       \\
     \% improvement &      8.89     &   6.20   &        3.62    \\
    \bottomrule
  \end{tabular}
  \vspace{0.1cm}
  \caption{Results of compression on fine-tuned GPT2 on works of Sir A.C. Doyle (checkpoint-7000)}
  \label{tab:table}
\end{table}

\subsection{Knowledge Distilled GPT2 on a domain corpora}

Knowledge distillation is a process of training a smaller machine learning model to replicate the behavior of a large complex model. There have been SOTA transformer models whose distilled version is made open source, examples include BERT and distilBERT. The goal is to transfer the knowledge from a large and complex teacher model to a small and compact student model to reduce the computing requirements.

We obtain the results obtained from distilled gpt2 in which the teacher model is gpt2-xl(1.5 B) and student model is gpt2(124 M) . We perform  knowledge distillation on books written by a specific author and perform compression on the works of the same author and other authors and compare it with gpt2 baselines. We train the model on a  learning rate of 5e-7 and obtain the best checkpoint for model evaluation.

Results validate that the distilled gpt2 performs better in compression against standard gpt2 baselines on the works of the same author hinting at the efficiency of fine tuning. It is also observed that distilled gpt2 performs better with works of other authors but not to the same level as that of the original author validating that there is some degree of style transfer during knowledge distillation.

\begin{table}[H]
  \centering
  \begin{tabular}{llll}
    \toprule
     Author &   W.B.Yeats(original) & Abraham Lincoln & Winston Churchill\\
    \midrule
     GZIP  ratio  &       2.28        &  2.26               &      2.09           \\
     GPT2+GZIP ratio   &       3.33        &    3.98           &    3.62             \\
     distilGPT2+GZIP ratio&      3.60       &   4.17            &         3.81        \\
     \% improvement &     8.29          &   4.97       &       5.38      \\
    \bottomrule
  \end{tabular}
  \vspace{0.1cm}
  \caption{Results of compression on distill GPT2 on works of W.B.Yeats (checkpoint-7500)}
  \label{tab:table}
\end{table}

\begin{table}[H]
  \centering
  \begin{tabular}{llll}
    \toprule
     Author &   A.C. Doyle(original) & Zane Gray & Rudyard Kipling\\
    \midrule
     GZIP  ratio  &       2.10       &  2.17               &     2.28          \\
     GPT2+GZIP ratio   &       3.68          &     3.47             &   3.15    \\
     distilGPT2+GZIP ratio&       4.01     &   3.70              &        3.33   \\
     \% improvement &      8.85        &    6.48          &         5.67    \\
    \bottomrule
  \end{tabular}
  \vspace{0.1cm}
  \caption{Results of compression on distil GPT2 on works of Sir Arthur Conan Doyle (checkpoint-9000)}
  \label{tab:table}
\end{table}

While traditional style transfer has been shown to improve compression in some cases, we encountered an instance where no significant improvement was observed when attempting to transfer an author's style. We distilled GPT-2 on the TinyShakespeare dataset and tested compression on the works of Zane Grey and Rudyard Kipling. The results indicate that the distinct differences in writing style between Shakespeare and these later-period authors likely contributed to the minimal improvement or deterioration (in the second case) in compression performance. This highlights the complexity of style transfer in relation to compression across diverse literary styles.

\begin{table}[H]
  \centering
  \begin{tabular}{lll}
    \toprule
     Author &   Zane Gray  & Rudyard Kipling\\
    \midrule
     GZIP  ratio  &     2.17         &      2.28      \\
     GPT2+GZIP ratio   &     3.47         &     3.15      \\
     distilGPT2+GZIP ratio&  3.48     &       3.10      \\
     \% improvement &   0.30   &             -1.45 \\
    \bottomrule
  \end{tabular}
  \vspace{0.1cm}
  \caption{Results of compression on distil GPT2 on Tiny Shakespeare (checkpoint-8500)}
  \label{tab:table}
\end{table}

\subsection{Compression across different LLMs}

We perform two lines of comparison here. Firstly, we measure how compression varies with size of the model. We compare the compression performance of gpt2 (124 M), gpt2-medium (335), gpt2-large (774 M), gpt2-xl(1.5 billion) on 8 books of 4 different authors against gzip baseline. From the results, it's very clear that the larger the model, the better the compression results which validates that LLMs with a huge number of parameters are better predictors. The biggest GPT2 variant improves gzip compression by 72\%.

\begin{table}[H]
  \centering
  \begin{tabular}{llllll}
    \toprule
    Author     & GZIP ratio   & GPT2+GZIP & GPT2-medium+GZIP & GPT2-large+GZIP  & GPT2-xl+GZIP\\
    \midrule
    Author 1 & 2.07  & 3.49   &3.65 & 3.73 & 3.79\\
    Author 2     & 2.20 & 3.23  & 3.37 & 3.45 &  3.54\\
    Author 3     & 2.18 &3.27    & 3.46 & 3.55   & 3.63\\
    Author 4     & 2.08    & 3.43  &  3.56&  3.65 &  3.72\\
    average      & 2.13   &  3.36  &   3.51  & 3.59 &  3.67      \\
    \bottomrule
  \end{tabular}
  \vspace{0.1cm}
  \caption{Results for compression on raw gpt2 variants}
  \label{tab:table}
\end{table}

Analysis on the time taken for compression by LLMs of different sizes indicates that the time taken increases monotonically with the parameter count of the Pre-trained transformer used while its compression performance increases in a non-linear fashion. The results also show that despite keeping the  number of characters for compression constant, the time taken varies for different files independent of the GZIP compression sizes.

\begin{figure}[h] % Use the figure environment to ensure correct placement
  \centering
  \begin{minipage}{0.5\textwidth} % Adjust the width as needed
   \centering
    \includegraphics[width=\textwidth]{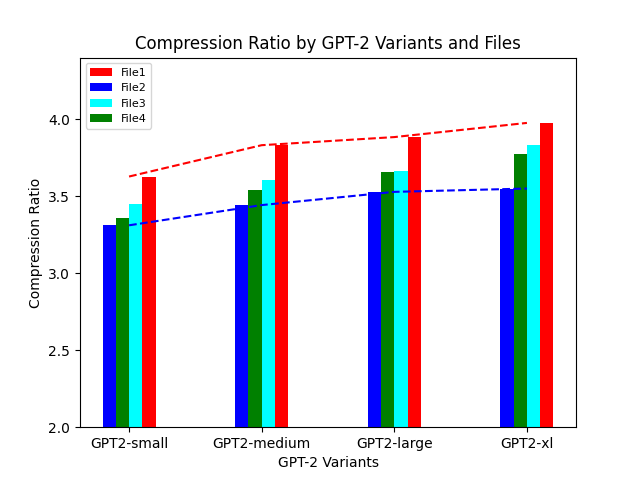} % Adjust the width as needed
    \caption{ Compression Ratio vs Model Size}
    \label{fig: Compression Ratio vs Model Size}
    
  \end{minipage}%
  \begin{minipage}{0.5\textwidth} % Adjust the width as needed
    \centering
    \includegraphics[width=\textwidth]{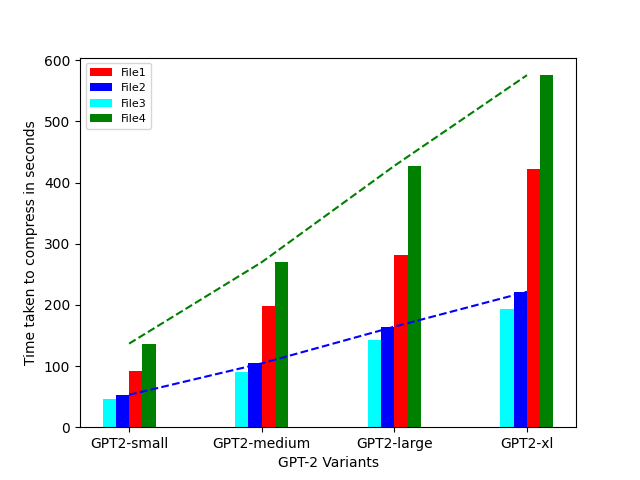} % Adjust the width as needed
    \caption{Time taken vs Model Size}
    \label{fig:Time taken vs Model Size}
  \end{minipage}
\end{figure}

\begin{table}[H]
\centering
 \begin{tabular}{lllllll}
      \toprule
      File &  Time  & Best Ratio & Best bpc & GZIP ratio &Entropy & Tokenized length\\
      \midrule
      File 1   & 421.42   & 3.97 & 2.01 & 2.06 &3.09  & 2461\\

      File 2   & 221.45   & 3.55 & 2.25 & 2.02 &3.10  &2733\\

      File 3   & 192.52  & 3.83 & 2.09 &  2.10 & 3.06 &2646\\

      File 4   & 575.58   & 3.77 &  2.12 & 2.26  & 2.99 &4339\\

      \bottomrule
    \end{tabular}
    \vspace{0.2cm}
    
    \caption{Data corresponding to the graphical representation}
    \label{tab:your_table}
\end{table}

\subsection{Compression Performance with Brotli}

With Brotli \cite{wikipedia_brotli} compression algorithm the compression ratio improved significantly, this is because Brotli utilises a dynamic dictionary based algorithm and works adaptively with the input. It outperforms Gzip's compression ratio. 

To evaluate the performance of Brotli we compressed first 1 lakh characters of Alice in Wonderland from the Gutenberg corpus. In the next experiment we distilled the gpt2-xl (1.5 billion)  model and transferred its knowledge to GPT2 (124 million) trained on a part of the Alice in Wonderland book for 50 epochs.  We tested compression performance on gpt2 (raw), gpt2-xl(teacher model) and distilled gpt2(student model).

\begin{table}[H]
  \centering
  \begin{tabular}{lllll}
    \toprule
     Length of input &   GZIP size (Bytes)  & Brotli size (Bytes) & GPT2+GZIP (Bytes) & GPT2+Brotli (Bytes)\\
    \midrule
     99900 & 37553 & 32645 & 34476 & 28876\\
     Ratio & 2.66 & 3.06 & 2.90 & 3.46 \\
    \bottomrule
  \end{tabular}
  \vspace{0.1cm}
  \caption{Results of compression with GPT2-xl on Alice in Wonderland}
  \label{tab:table}
\end{table}

Brotli has improved the compression ratio by 20\% from that of gzip in the given experiment. 

\begin{table}[H]
  \centering
  \begin{tabular}{llll}
   Length of input : 28851     GZIP size: 11568 \\
    \toprule
     Model  & GPT2 & GPT2-xl & distilled GPT2 \\
    \midrule
     Size with GPT2+GZIP& 11048& 10150   & 7151\\
     Ratio with GPT2+GZIP & 2.61 & 2.84 & 4.03 \\
     Size with GPT2+Brotli &  9104  & 8369 &  6203 \\
     Ratio with GPT2+Brotli & 3.16 &   3.44     &   4.65   \\
    \bottomrule
  \end{tabular}
  \vspace{0.1cm}
  \caption{Results of compression with distil GPT2 (checkpoint 16000) on Alice in Wonderland}
  \label{tab:table}
\end{table}

This enhances the efficiency of the process by not only improving accuracy but also reducing time complexity, as the distilled model is a less dense neural network architecture. The student model achieves a 47\% better compression rate compared to the untrained student model and outperforms the teacher model by 35\%.

\subsection{Compression on multi-lingual text}

In this experiment, we compressed French and Hindi text from the OPUS Corpora \footnote{\href{https://opus.nlpl.eu/}{OPUS - open parallel corpora}}. The input size was 100,000 characters with a context length of 100 characters. For French text, raw GPT-2 showed minimal improvement over the gzip baseline. However, with Hindi text, predictive compression using GPT-2 failed, resulting in expansion rather than compression. This performance is likely due to two factors: the tokenized length of Hindi characters is longer than French or English, and GPT-2's training data was predominantly English. The results are summarized in the following table.

\begin{table}[H]
  \centering
  \begin{tabular}{llll}
    \toprule
     Model  & English & French& Hindi\\
    \midrule
     Ratio with GZIP  &  2.48   &    2.51 &    2.17   \\
     Ratio with GPT2+GZIP &    2.95   & 2.69 &    0.36\\
     Ratio with GPT2+Brotli &   3.49    & 3.18 &   0.40   \\
    \bottomrule
  \end{tabular}
  \vspace{0.1cm}
  \caption{Results of compression on French and Hindi text using raw GPT2}
  \label{tab:table}
\end{table}

We further tested the compression capabilities of fine-tuned GPT-2 models on French \footnote {\href{https://huggingface.co/dbddv01/gpt2-french-small}{Fine-tuned model on French text}} and Hindi \footnote {\href{https://huggingface.co/surajp/gpt2-hindi}{Fine-tuned model on Hindi text}} text. The results showed improved compression performance, supporting the claim that domain-specific compression outperforms the vanilla GPT-2. The improvement was more significant for Hindi text, likely due to the greater syntactic differences between Hindi and English compared to the closer similarities between French and English.

\begin{table}[H]
  \centering
  \begin{tabular}{lll}
    \toprule
     Model  & French& Hindi\\
    \midrule
     Ratio with GZIP  &    2.51 &    2.17   \\
     Ratio with Brotli  &      2.80 &    2.61   \\
     Ratio with Fine-tuned GPT2+GZIP & 3.74 &  3.07  \\
     Ratio with Fine-tuned GPT2+Brotli & 4.51 &   3.37   \\
     Percentage improvement (Brotli) &   41.8\%    &    742.5\%       \\
    \bottomrule
  \end{tabular}
  \vspace{0.1cm}
  \caption{Results of compression on French and Hindi text using Fine-tuned GPT2}
  \label{tab:table}
\end{table}

\section{Inferences}

\begin{enumerate}
\item Compression ratio improves by at least 57\% in neural networks based predictive compression method when compared against the GZIP standards.
\item On fine-tuning and Knowledge distillation compression performance improves by at least 10\%  upto 10000 runs. With a higher amount of training data and a larger number of epochs improvement close to 50\% is expected as can be seen in the Alice in Wonderland case.
\item When multi-lingual text are analysed, vanilla GPT2 performs better compression in English text when compared to French and Hindi text. Fine-tuning improves compression performance on French and Hindi text. 
\item Compression performance depends on the size of the input text, on average neural compression compresses the original file size by 3.6x using the biggest model GPT2-xl.
\item Brotli compression mechanism is more efficient than GZIP mechanism and improves the neural compression by nearly 20\%. 
\item Time complexity of neural compression process is roughly proportional to the model size. Compression ratio increases monotonically with model size.
\item Compression ratio of neural compression method does not solely depends on the size of the file. It is an interplay of entropy, tokenized length and the content of the file.
\end{enumerate}

\section{Conclusion and Future Work}

In conclusion, our experiments demonstrate that rank-based predictive compression offers a more effective approach compared to traditional information-theoretic methods, particularly when leveraging context through fine-tuned or knowledge-distilled models. These techniques show promise in achieving better compression ratios, highlighting the potential of predictive compression in diverse applications. While our study focused on a subset of open-source large language models, future work could expand this scope by incorporating models (such as LLaMa and Zephyr), and further comparing them with the GPT-2 family. Future research can also include developing suitable information theoretic compressors that complement the neurally predicted ranks in reducing redundancies.  As advancements in compression algorithms like Brotli and Burrows-Wheeler align with ongoing developments in neural networks, predictive compression stands poised for substantial growth. The continuous evolution of machine learning techniques presents opportunities for refining and optimizing compression strategies, marking this as a dynamic area for future research.

\newpage
%Bibliography
\bibliographystyle{unsrt}  
\bibliography{references}  
\

\end{document}